# Oligoyne molecular junctions for efficient room temperature thermoelectric power generation

Hatef Sadeghi[*], Sara Sangtarash, and Colin J Lambert[*]

Quantum Technology Centre, Lancaster University, LA1 4YB Lancaster, UK
h.sadeghi@lancaster.ac.uk, c.lambert@lancaster.ac.uk

*Abstract*

Understanding phonon transport at a molecular scale is fundamental to the development of high-performance thermoelectric materials for the conversion of waste heat into electricity. We have studied phonon and electron transport in alkane and oligoyne chains of various lengths and find that due to the more rigid nature of the latter, the phonon thermal conductances of oligoynes are counter intuitively lower than that of the corresponding alkanes. The thermal conductance of oligoynes decreases monotonically with increasing length, whereas the thermal conductance of alkanes initially increases with length and then decreases. This difference in behaviour arises from phonon filtering by the gold electrodes and disappears when higher-Debye-frequency electrodes are used. Consequently a molecule that better transmits higher-frequency phonon modes, combined with a low-Debye-frequency electrode that filters high-energy phonons is a viable strategy for suppressing phonon transmission through the molecular junctions. The low thermal conductance of oligoynes, combined with their higher thermopower and higher electrical conductance lead to yield a maximum thermoelectric figure of merit of $ZT$ = 1.4, which is several orders of magnitude higher than for alkanes.

For many years, the attraction of the single-molecule electronics [1-6] has stemmed from their potential for sub-10nm electronic switches and rectifiers, and from their provision of sensitive platforms for single-molecule sensing. In the recent years, their potential for removing heat from nanoelectronic devices (thermal management) and thermoelectrically converting waste heat into electricity [7-10] has also been recognised. The efficiency of a thermoelectric device for power generation is characterised by the dimensionless figure of merit $ZT = GS^2T/\kappa$, where $G$ is the electrical conductance, $S$ is the thermopower (Seebeck coefficient), $T$ is temperature and $\kappa$ is the thermal conductance [11-14]. Therefore low-$\kappa$ materials are needed for efficient conversion of heat into electricity, whereas materials with high $\kappa$ are needed for thermal management. Inorganic materials for thermoelectricity have been extensively studied and have delivered $ZT$ values as high as 2.2 at temperatures over 900K [15]. However this level of efficiency does not meet the requirements of current energy demands [16] and furthermore, the materials are difficult to process and have limited global supply. Organic thermoelectric materials may be an attractive alternative, but at present the best organic thermoelectric material with a $ZT$ of 0.6 in room temperature [17,18] is still not competitive with inorganics. In an effort to overcome these limitations, single organic molecules and self-assembled monolayers have attracted recent scientific interest, both for their potential as room temperature thermoelectric materials and thermal management [19,20].

Strategies for reducing the denominator (ie $\kappa$) of ZT in single-molecule junctions are fundamentally different from inorganic bulk materials. In the latter, phonon transport can be reduced by nanostructuring [21,22], whereas molecular junctions are naturally nanostructured and additional strategies based on molecular phonon conversion [23] become possible, including the reduction of thermal conductance due to weak overlap between the continuum of vibrational states in the electrodes and discrete vibrational states of the molecules or the weak interaction between different part of the molecules, as in π-π stacked structures [24]. On the other hand, strategies for increasing the numerator of ZT (ie the power factor) focus on tuning electron transport properties, which are determined by the energetic position of the electrode Fermi energy relative to transport resonances through the frontier orbitals of the molecule. If $T_{el}(E)$ is the transmission coefficient of electrons of energy E passing from one electrode to the other through a molecule, then the thermopower S is approximately proportional to the slope of the $ln\ T_{el}(E)$, evaluated at the Fermi energy $E_F$, whereas the electrical conductance is proportional to $T_{el}(E_F)$. Therefore if the Fermi energy lies in a region of high slope, close to a transmission resonance of the frontier orbitals and provided the (HOMO-LUMO) gap between the resonances is greater than ~ $4k_BT$ (ie ~ *100 meV* at room temperature), then both G and S can be enhanced [11]. In the literature, there are many experiments addressing electronic properties of single molecules, but far fewer addressing single-molecule phonon transport, partly because it is extremely difficult to measure the thermal conductance of a single molecule. This difficulty is partly circumvented by scanning thermal microscope measurements of a few thousands of molecules in parallel, such as a recent experimental study of the length-dependent thermal conductance of alkanes by the IBM group [25], which revealed a surprising initial increase in thermal conductance with length for short alkanes.

In this letter, we present a comparative theoretical study of the length dependent thermal properties of the alkanes and oligoynes, which elucidates the origin of this initial increase and demonstrate that oligoynes offer superior performance for future efficient thermoelectric power generation. Since the thermopower and electrical conductance of oligoynes and alkanes are generally understood [26-28] our main focus in this letter is to calculate the thermal conductance of these materials, which contains contributions from both electrons and phonons. The main unexpected result from our study is that thermal conductances of oligoynes are lower than alkanes of the same length, which is counter intuitive, because alkanes are more floppy than oligoynes. Moreover, the thermopower and electrical conductance is higher in oligoynes. The resulting combination of low thermal conductance, high thermopower and high electrical conductance lead to a high value of ZT and make oligoynes attractive for future thermoelectric devices.

Recently length-dependent thermopowers of alkane, alkene, and oligoyne chains with four different anchor groups (thiol, isocyanide, and amine end groups and direct coupling) were theoretically studied for the chains with the length 2, 4, 6 and 8 carbon atoms [26-28]. It was shown that the sign and magnitude of the thermopower, and the conductance-length attenuation factor ($\beta$) are strongly affected by the anchor groups. For example, in oligoynes the thermopower was found to be positive with a direct C-Au bond or thiol anchor and negative with a NC end group [29]. Furthermore, while the conductance G decays exponentially as $exp(-\beta L)$ with increasing molecular length L, the thermopower shows a linear length dependence [30]. The crucial point is that higher thermopower is predicted for oligoynes compared with alkanes for all lengths [29], a fact that is in good agreement with our calculations below. Although the thermal conductance of self-assembled monolayers of alkanes sandwiched between gold and GaAs was shown to be length independent and as high as 27 *MW* $m^{-2}K^{-1}$ [23], recent experiment on alkanes sandwiched between gold and $SiO_2$ shows length dependencies [25] in agreement with our study in this letter.

As shown in figures 1a and 2a, the alkanes and oligoynes of interest in this study are of lengths of 2, 4, 8 and 16 carbon atoms ($N = 1, 2, 4,$ and $8$) and are connected to two gold electrodes through dihydrobenzo[b]thiophene (BT) anchor groups. To study the thermal properties of the alkanes (fig. 1a) and oligoynes (fig. 2a), we use density functional theory (*DFT*) to calculate their electronic and vibrational properties within the junction. We first carry out geometry optimization of each molecule placed between two gold electrodes using *DFT* [31,32] to find the ground state optimum positions of the atoms (*q*) relative to each other's and electronic mean field Hamiltonian of the system including electrodes and molecule (see methods). The mean field Hamiltonian is combined with our Green's function scattering method [33] to calculate the electron transmission coefficient $T_{el}(E)$, from which the electrical conductance $G = G_0 \int dE\, T_{el}(E_F)(-\partial f/\partial E)$ is obtained, where $G_0$ is the conductance quantum and $f(E)$ is the Fermi function. Figures 1b and 2b show $T_{el}(E)$ for alkanes with $N = 1, 2, 4$ and $8$ pairs of carbon atoms and oligoynes with $N = 1, 2$ and $4$ pairs of carbon atoms, respectively. In agreement with previous experimental and theoretical studies [26,34-37], when evaluated at the *DFT*-computed Fermi energy ($E_F^{DFT}$) the transmission coefficients decrease with the length for both alkanes and oligoynes. However, the conductances of the oligoynes are higher than those of the alkanes for equivalent lengths, due to the broken π-conjugation in the alkanes.

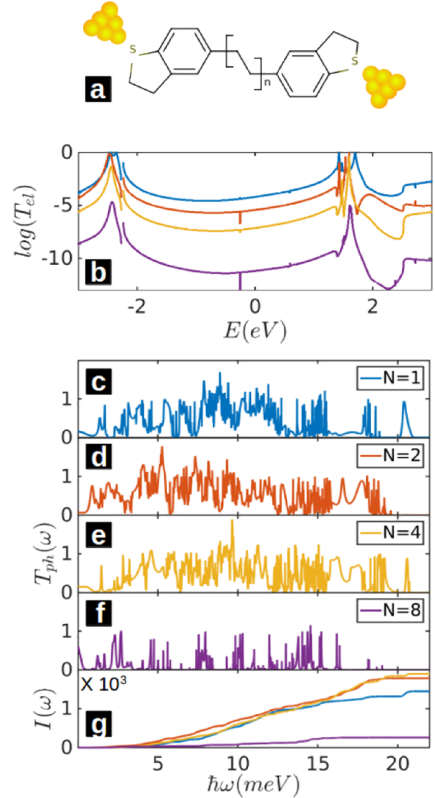

Figure 1: Molecular structure and transport properties of alkanes with the length of $N = 1, 2, 4, 8$ between two gold electrodes. (a) Schematic of the junction, (b) electronic and (c-f) phononic transmission coefficients. (g) The integrated phonon transmission $I(\omega)$. All molecules are terminated with BT (ie dihydrobenzo[b]thiophene) anchor groups.

To calculate the vibrational modes of each structure, we use the harmonic approximation method to construct the dynamical matrix *D*. Each atom is displaced from its equilibrium position by δ*q'* and –δ*q'* in *x*, *y* and *z* directions and the forces on all atoms calculated in each case. For *3n* degrees of freedom (*n* = number of atoms), the *3n × 3n* dynamical matrix $D_{ij} = (F_i^q(\delta q_j') - F_j^q(-\delta q_j'))/2M_{ij}\delta q_j'$ is constructed, where *F* and *M* are the force and mass matrices (see methods). For an isolated molecule, the square root of the eigenvalues of *D* determine the frequencies ω associated with the vibrational modes of the molecule in the junction. For a molecule within a junction, the dynamical matrix describes an open system composed of the molecule and two semi-infinite electrodes and is used to calculate the phononic transmission coefficient $T_{ph}(\omega)$ for the phonons with energy $\hbar\omega$ passing through the molecule from the right to the left electrode. Figures 1c-f and figure 2c-e show $T_{ph}(\omega)$ for alkanes and oligoynes respectively of different lengths. To elucidate the different areas under these curves, figure 1g and 2f show their integrated transmission coefficients $I(\omega) = \int_0^\omega T_{ph}(\omega)d\omega$. At high frequencies, the transmission is limited by the number of open phonon channels in the gold electrodes (see fig. S5 in Supplementary Information *SI*), which falls to zero above the gold (111) Debye frequency of 21 *meV*. This shows that a molecule that better transmits higher modes, combined with a low-Debye-frequency electrode that filters high energy phonons could be a viable strategy to suppress phonon transmission through the junction.

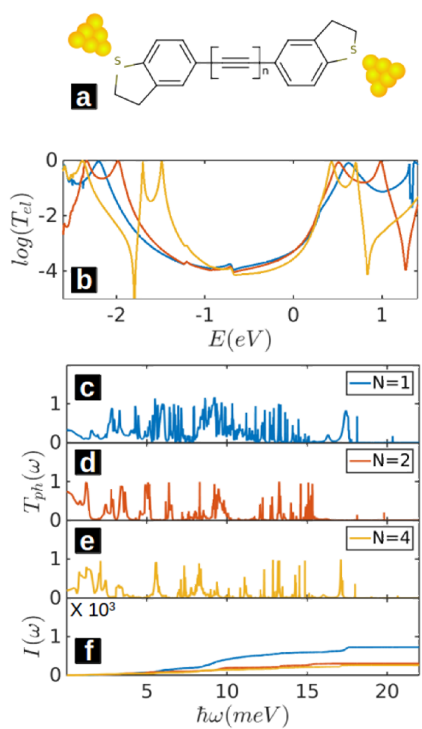

Figure 2: Molecular structure and transport properties of oligoynes with the length of *N* = 1, 2, 4 between two gold electrodes. (a) Schematic of the junction, (b) electronic and (c-e) phononic transmission coefficients. (f) The integrated phonon transmission *I(ω)*. All molecules are terminated with BT (ie dihydrobenzo[b]thiophene ) anchor groups.

The thermal conductance of the junction ($\kappa = \kappa_{ph} + \kappa_{el}$) is obtained by summing the contributions from both electrons ($\kappa_{el}$) and phonons ($\kappa_{ph}$). The electronic (phononic) thermal conductances are calculated from the electronic (phononic) transmission coefficients shown in figures 1b and 2b (figs. 1c-f and 2c-e) as described in the methods section below. Figure 3a,b shows the resulting electronic thermal conductance $\kappa_{el}$ and figure 3c,d the phononic thermal conductance $\kappa_{ph}$ for alkanes and oligoynes, respectively. In general, the thermal conductance is dominated by phonons. For example the phononic thermal conductance $\kappa_{ph}$ of the *N* = 1 oligoyne (alkane) is more than 30 (700) times bigger than electronic thermal conductance $\kappa_{el}$ at room temperature. Therefore for these molecules $\kappa_{el}$ is negligible. In addition, the thermal conductances of the alkanes are higher than those of the oligoynes, which suggests that alkanes are potentially useful for thermal management, but less useful for thermoelectricity.

In both types of junctions, the thermal conductance increases with temperature up to about 170K, (ie the Debye temperature of the gold electrodes) and then remains constant (fig. 3c,d). For oligoynes, at room temperature, $\kappa$ is equal to 15.6, 9.2 and 7.7 *pW/K* for *N* = 1, 2, 4 respectively. It is apparent that the thermal conductance of the oligoyne molecules decreases monotonically with length. In contrast, for alkanes, thermal conductance initially increases with length and then decreases. At room temperature, the $\kappa$ of alkanes are equal to 25.4, 33.4, 30.3 and 5.6 *pW/K* for $N = 1, 2, 4$ and 8 respectively, revealing that the $N = 2$ alkane has the highest thermal conductance. This initial increase in thermal conductance with length followed by a decrease has been observed experimentally [25] and predicted theoretically [38] in previous studies, although its origin remains unexplored. To account for this behaviour it is useful to understand why it does not occur for the more rigid oligoynes.

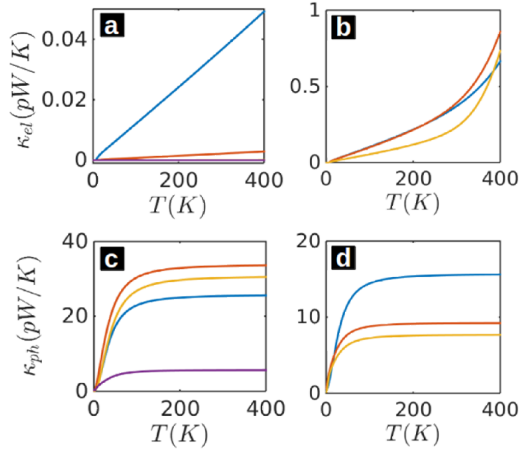

Figure 3: Electronic and phononic thermal conductance of alkanes and oligoynes. (a,b) show the electronic thermal conductance of alkanes and oligoynes respectively. (c,d) show the phononic thermal conductance of alkanes and oligoynes respectively. Results are shown for molecules of different lengths. For colour coding see figure 1.

First we note that the phonon thermal conductance of oligoynes is lower than that of alkanes, because due to their rigidity, the phonon level spacing between the oligoyne modes is bigger than that of the alkanes (fig. 1c-f and fig. 2c-e). Consequently a greater fraction of the oligoyne modes lie above the Debye frequency of the gold electrodes and are therefore filtered by the gold. In the less-rigid alkanes, which possess more low-frequency modes (fig. 1c-f), this filtering effect is less pronounced. As the length of the chain is increased, all modes move to lower frequencies (which tends to increase the thermal conductance) and the widths of transmission resonances decrease (which tends to decrease the thermal conductance), because the imaginary part of the self-energy is proportional to the inverse length of the molecules [39].

This unconventional behaviour is also illustrated by a simple "tight-binding" model (see fig. S3 in *SI*) with one degree of freedom per atom. In oligoynes, relatively-high frequency of the modes means that resonance narrowing dominates at all lengths. In the case of alkanes, a significant number of the $N=1$ modes are filtered by the gold and upon increasing to $N=2$, these modes move to lower frequencies and are no longer filtered, leading to the unexpected increase in thermal conductance. At longer lengths decrease in resonance widths with increasing length dominates and $\kappa_{ph}$ decreases with length. To demonstrate that this counter-intuitive effect disappears when phonon filtering is removed, we have examined the effect of artificially reducing the mass of the electrodes gold atoms. This is achieved by simply multiplying the mass matrix $M_{ij}$ by a scale factor in the *DFT*-constructed dynamical matrix, which increases the Debye frequency of the electrodes. As shown in figure S4 in *SI*, the resulting thermal conductance of the alkanes decreases monotonically with length. This leads us to predict that conventional length dependence for thermal conductance of alkanes will be observed if higher-Debye-frequency electrodes such as graphene[5] are used.

In addition, to demonstrate the effect of the junction properties e.g. different anchor and electrode surface on the phononic thermal conductance, we have examined the effect of artificially changing the mass of the atoms in the BT anchors by a factor of 0.5 or 2 and of the atoms on the surface of the electrodes by a factor of 2 as shown in figure S10 in *SI*. Although thermal conductance is affected by changes in the anchor or electrodes surface, in all cases, the alkanes show an initial rise in $\kappa_{ph}$ upon increasing the length from *N*=1 to *N*=2 and the thermal conductance of alkanes is higher than that of the corresponding oligoynes. Furthermore, to demonstrate that this trend is resilient and independent of the anchor and electrode surface configuration, using *DFT*-constructed dynamical matrix, we have calculated the phononic thermal conductance of the alkane and oligoyne with the length of *N*=4 with two other anchors (amine and thiol anchors) and a different configuration of the electrode tip as shown in figure S6 in *SI*. The corresponding electronic and phononic transmission coefficients are shown in figure S7 and S8 for alkanes and oligoynes, respectively. Although the amplitude of the thermal conductance varies with anchor and tip configuration (fig. S9 in *SI*), the thermal conductance of oligoynes is still lower than that of the corresponding alkanes.

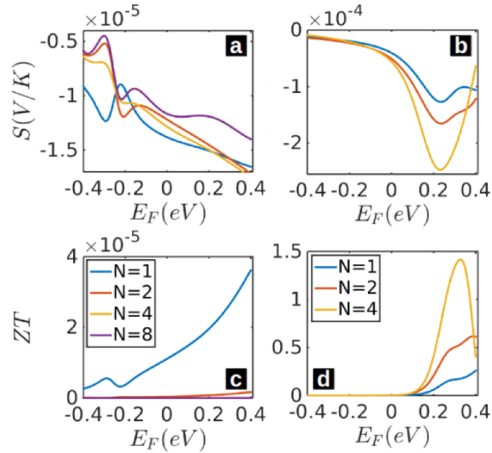

Figure 4: Thermopower *S* and full thermoelectric figure of merit *ZT* for alkanes and oligoynes. (a,b) show the thermopower of alkanes and oligoynes respectively. (c,d) show the figure of merit of alkanes and oligoynes respectively. Results are shown for molecules of different lengths *N*.

To compare the potential of these two families of molecules for thermoelectricity, we calculated the electrical conductance and thermopower of the alkanes and oligoynes as described in the methods section. Since the thermopower depends on the Fermi energy of the leads and could be tuned by electrostatic or electrochemical gating or doping, we computed the electrical conductance, thermopower and total thermoelectric figure of merit at different Fermi energies. Figures 4a,b show the thermopower of alkanes and oligoynes respectively with different lengths at room temperature and for different Fermi energies. In general the thermopower is an order of magnitude higher for oligoynes. This is because for oligoynes, the Fermi energy lies in the tail of the HOMO or LUMO resonance [26] depending on the anchor groups (see eg fig. 2b and fig. S8 in *SI*) and therefore the slope of the *ln* $T_{el}(E)$ is high, which leads to higher thermopower. In contrast, the Fermi energy for alkanes is near the middle of the HOMO-LUMO gap (see fig. 1b), where the slope of the *ln* $T_{el}(E)$ is much lower. The electrical conductance of alkanes also is lower than oligoynes due to the broken π-conjugation (fig. 1b and 2b). Simultaneously, as discussed above, the thermal conductance is lower for oligoynes. Combining the high electrical conductance of oligoynes, with their high thermopower and low thermal conductance, yields a maximum *ZT* of 1.4 (at a Fermi energy of 0.3*eV*), which is several orders of magnitude higher than for alkanes, as shown in figures 4c,d. To achieve this high value, the Fermi energy should be optimally located in the tail of the LUMO resonance, which could be achieved by doping, or gating the molecules.

In conclusion, understanding phonon and electron transport through molecules attached to metallic electrodes is crucial to the development of high-performance thermoelectric materials and to thermal management in nanoscale devices. We have studied simultaneously phonon and electron transport in alkane and oligoyne chains as model systems and find that due to the more rigid nature of the latter, the phonon thermal conductances of oligoynes are lower than that of the corresponding alkanes. Therefore in view of their higher thermal conductance, we conclude that alkanes are the better candidates for thermal management. The thermal conductance of oligoynes decreases monotonically with increasing length, whereas the thermal conductance of alkanes initially increases with length and then decreases. This difference in behaviour arises from phonon filtering by the gold electrodes and leads us to predict that the initial rise in thermal conductance of alkanes would disappear if higher-Debye-frequency electrodes such as graphene are used. Furthermore, by comparing results for different anchor groups and tip configurations we conclude that the above trends are resilient. This is a significant result, because it demonstrates that not just the molecule alone, but combinations of molecules and electrodes and their interplay should be included in design strategies for future organic-molecule-based thermoelectricity. The low thermal conductance of oligoynes, combined with their higher thermopower and higher electrical conductance yield a maximum *ZT* of 1.4, which is several orders of magnitude higher than for alkanes. Therefore oligoynes are attractive candidates for high-performance thermoelectric energy conversion.

**Methods**

The geometry of each structure consisting of the gold electrodes and a single molecule (alkane or oligoyne) was relaxed to the force tolerance of 20 meV/Å using the *SIESTA*[32] implementation of density functional theory (DFT), with a double-ζ polarized basis set (DZP) and the Generalized Gradient Approximation (GGA) functional with Perdew-Burke-Ernzerhof (PBE) parameterization. A real-space grid was defined with an equivalent energy cut-off of 250 Ry. From the relaxed *xyz* coordinate of the system, sets of *xyz* coordinates were generated by displacing each atom in positive and negative *x*, *y* and *z* directions by $\delta q' = 0.01$ Å. The forces in three directions $q_i = (x_i, y_i, z_i)$ on each atom were then calculated by DFT without geometry relaxation. These sets of the force $F_i^q = (F_i^x, F_i^y, F_i^z)$ are used to construct the dynamical matrix as:

$$D_{ij} = \frac{K_{ij}^{qq'}}{M_{ij}} \quad (1)$$

where $K_{ij}^{qq'}$ for $i \neq j$ are obtained from finite differences

$$K_{ij}^{qq'} = \frac{F_i^q(\delta q_j') - F_j^q(-\delta q_j')}{2\delta q_j'} \quad (2)$$

and the mass matrix $M = \sqrt{M_i M_j}$. To satisfy momentum conservation, the *K* for $i = j$ (diagonal terms) is calculated from $K_{ii} = -\sum_{i \neq j} K_{ij}$. The phonon transmission $T_{ph}(\omega)$ then can be calculated from the relation:

$$T_{ph}(\omega) = Tr(\Gamma_L^{ph}(\omega) G_{ph}^R(\omega) \Gamma_R^{ph}(\omega) G_{ph}^{R\dagger}(\omega)) \quad (3)$$

In this expression, $\Gamma_{L,R}^{ph}(\omega) = i\left(\Sigma_{L,R}^{ph}(\omega) - \Sigma_{L,R}^{ph\,\dagger}(\omega)\right)$ describes the level broadening due to the coupling between left ($L$) and right ($R$) electrodes and the central scattering region formed from the molecule and closest contact layers of gold, $\Sigma_{L,R}^{ph}(\omega)$ are the retarded self-frequencies associated with this coupling and $G_{ph}^R = \left(\omega^2 I - D - \Sigma_L^{ph} - \Sigma_R^{ph}\right)^{-1}$ is the retarded Green's function, where $D$ and $I$ are the dynamical and the unit matrices, respectively. The phonon thermal conductance $\kappa_{ph}$ at temperature $T$ is then calculated from:

$$\kappa_{ph}(T) = \frac{1}{2\pi} \int_0^\infty \hbar\omega T_{ph}(\omega) \frac{\partial f_{BE}(\omega, T)}{\partial T} d\omega \qquad (4)$$

where $f_{BE}(\omega, T) = (e^{\hbar\omega/k_B T} - 1)^{-1}$ is Bose–Einstein distribution function and $\hbar$ is reduced Planck's constant and $k_B = 8.6 \times 10^{-5}\ eV/K$ is Boltzmann's constant.

To calculate electronic properties of the molecules in the junction, from the converged DFT calculation, the underlying mean-field Hamiltonian $H$ was combined with our quantum transport code, *GOLLUM* [33]. This yields the transmission coefficient $T_{el}(E)$ for electrons of energy $E$ (passing from the source to the drain) via the relation $T_{el}(E) = Tr(\Gamma_L^{el}(E) G_{el}^R(E) \Gamma_R^{el}(E) G_{el}^{R\dagger}(E))$ where $\Gamma_{L,R}^{el}(E) = i\left(\Sigma_{L,R}^{el}(E) - \Sigma_{L,R}^{el\,\dagger}(E)\right)$ describes the level broadening due to the coupling between left ($L$) and right ($R$) electrodes and the central scattering region, $\Gamma_{L,R}^{el}(E)$ are the retarded self-energies associated with this coupling and $G_{el}^R = \left(ES - H - \Sigma_L^{el} - \Sigma_R^{el}\right)^{-1}$ is the retarded Green's function, where $H$ is the Hamiltonian and $S$ is the overlap matrix obtained from *SIESTA*. Using the approach explained in [11,14,33], the electrical conductance $G_{el}(T) = G_0 L_0$, the electronic contribution of the thermal conductance $\kappa_e(T) = (L_0 L_2 - L_1^2)/hTL_0$ and the thermopower $S(T) = -L_1/eTL_0$ of the junction are calculated from the electron transmission coefficient $T_{el}(E)$ where:

$$L_n(T) = \int_{-\infty}^{+\infty} dE\ (E - E_F)^n\ T_{el}(E) \left(-\frac{\partial f_{FD}(E, T)}{\partial E}\right) \qquad (5)$$

and $f_{FD}(E, T)$ is the Fermi-Dirac probability distribution function $f_{FD}(E, T) = (e^{(E-E_F)/k_B T} + 1)^{-1}$, $T$ is the temperature, $E_F$ is the Fermi energy, $G_0 = 2e^2/h$ is the conductance quantum, $e$ is electron charge and $h$ is the Planck's constant. Since the above methodology ignores phonon-phonon and electron-phonon scattering, our agreement with the measurements of ref [25] suggests that such inelastic scattering is not a large effect. This is consistent with measurements on other molecules, which suggest that inelastic scattering of electrons at room temperature is a small effect, provided the length of the molecule is less than approximately $3\ nm$ (see eg [40,41]). In our calculations, the size of the molecules are between 1.2 to 2.9 $nm$.

**Acknowledgment**

This work is supported by UK EPSRC grants EP/K001507/1, EP/J014753/1, EP/H035818/1 and the European Union Marie-Curie Network MOLESCO.

**References**

(1) Lambert, C. *Chemical Society Reviews* **2015**, 875.

(2) Aradhya, S. V.; Venkataraman, L. *Nat. Nano.* **2013**, *8*, 399.

(3) Geng, Y.; Sangtarash, S.; Huang, C.; Sadeghi, H.; Fu, Y.; Hong, W.; Wandlowski, T.; Decurtins, S.; Lambert, C. J.; Liu, S.-X. *J. Am. Chem. Soc.* **2015**, *137*, 4469.

(4) Widawsky, J. R.; Chen, W.; Vázquez, H.; Kim, T.; Breslow, R.; Hybertsen, M. S.; Venkataraman, L. *Nano Lett.* **2013**, *13*, 2889.

(5) Sadeghi, H.; Mol, J. A.; Lau, C. S.; Briggs, G. A. D.; Warner, J.; Lambert, C. J. *Proc. Natl. Acad. Sci.* **2015**, *112*, 2658.

(6) Sun, L.; Diaz-Fernandez, Y. A.; Gschneidtner, T. A.; Westerlund, F.; Lara-Avila, S.; Moth-Poulsen, K. *Chemical Society Reviews* **2014**, *43*, 7378.

(7) Moore, A. L.; Shi, L. *Materials Today* **2014**, *17*, 163.

(8) Kim, Y.; Jeong, W.; Kim, K.; Lee, W.; Reddy, P. *Nat. Nanotechnol.* **2014**, *9*, 881.

(9) Widawsky, J. R.; Darancet, P.; Neaton, J. B.; Venkataraman, L. *Nano Lett.* **2011**, *12*, 354.

(10) Sadeghi, H.; Sangtarash, S.; Lambert, C. J. *Beilstein Journal of Nanotechnology* **2015**, *6*, 1413.

(11) Sadeghi, H.; Sangtarash, S.; Lambert, C. J. *Sci. Rep.* **2015**, *5*, 9514.

(12) Reddy, P.; Jang, S.-Y.; Segalman, R. A.; Majumdar, A. *Science* **2007**, *315*, 1568.

(13) Zimbovskaya, N. A. *J. Phys.: Condens. Mat.* **2014**, *26*, 275303.

(14) Sadeghi, H.; Sangtarash, S.; Lambert, C. J. *Beilstein Journal of Nanotechnology* **2015**, *6*, 1176.

(15) Zhao, L.-D.; Lo, S.-H.; Zhang, Y.; Sun, H.; Tan, G.; Uher, C.; Wolverton, C.; Dravid, V. P.; Kanatzidis, M. G. *Nature* **2014**, *508*, 373.

(16) Snyder, G. J.; Toberer, E. S. *Nature materials* **2008**, *7*, 105.

(17) Chung, D.-Y.; Hogan, T.; Brazis, P.; Rocci-Lane, M.; Kannewurf, C.; Bastea, M.; Uher, C.; Kanatzidis, M. G. *Science* **2000**, *287*, 1024.

(18) Zhao, L.-D.; Dravid, V. P.; Kanatzidis, M. G. *Energy & Environmental Science* **2014**, *7*, 251.

(19) Zhang, Q.; Sun, Y.; Xu, W.; Zhu, D. *Advanced Materials* **2014**, *26*, 6829.

(20) Majumdar, S.; Sierra-Suarez, J. A.; Schiffres, S. N.; Ong, W.-L.; Higgs, I., C Fred; McGaughey, A. J.; Malen, J. A. *Nano Lett.* **2015**.

(21) Fagas, G.; Kozorezov, A.; Lambert, C.; Wigmore, J.; Peacock, A.; Poelaert, A.; den Hartog, R. *Phys. Rev. B* **1999**, *60*, 6459.

(22) Kambili, A.; Fagas, G.; Fal'ko, V.; Lambert, C. *Phys. Rev. B* **1999**, *60*, 15593.

(23) Wang, R. Y.; Segalman, R. A.; Majumdar, A. *Appl. Phys. Lett.* **2006**, *89*, 173113.

(24) Kiršanskas, G.; Li, Q.; Flensberg, K.; Solomon, G. C.; Leijnse, M. *Appl. Phys. Lett.* **2014**, *105*, 233102.

(25) Meier, T.; Menges, F.; Nirmalraj, P.; Hölscher, H.; Riel, H.; Gotsmann, B. *Phys. Rev. Lett.* **2014**, *113*, 060801.

(26) Hüser, F.; Solomon, G. C. *J. Phys. Chem. C* **2015**, *119*, 14056.

(27) Tan, A.; Balachandran, J.; Sadat, S.; Gavini, V.; Dunietz, B. D.; Jang, S.-Y.; Reddy, P. *J. Am. Chem. Soc.* **2011**, *133*, 8838.

(28) Baheti, K.; Malen, J. A.; Doak, P.; Reddy, P.; Jang, S.-Y.; Tilley, T. D.; Majumdar, A.; Segalman, R. A. *Nano Lett.* **2008**, *8*, 715.

(29) Karlström, O.; Strange, M.; Solomon, G. C. *J. Chem. Phys* **2014**, *140*, 044315.

(30) Pauly, F.; Viljas, J. K.; Cuevas, J. C. *Phys. Rev. B* **2008**, *78*, 035315.

(31) Soler, J. M.; Artacho, E.; Gale, J. D.; García, A.; Junquera, J.; Ordejón, P.; Sánchez-Portal, D. *J. Phys.: Condens. Mat.* **2002**, *14*, 2745.

(32) Perdew, J. P.; Burke, K.; Ernzerhof, M. *Phys. Rev. Lett.* **1996**, *77*, 3865.

(33) Ferrer, J.; Lambert, C.; Garcia-Suarez, V.; Zsolt Manrique, D.; Visontai, D.; Oroszlani, L.; Ferradás, R.; Grace, I.; Bailey, S.; Gillemot, K.; Sadeghi, H.; Algharagholy, L. *New J Phys* **2014**, *16*, 093029.

(34) Moreno-García, P.; Gulcur, M.; Manrique, D. Z.; Pope, T.; Hong, W.; Kaliginedi, V.; Huang, C.; Batsanov, A. S.; Bryce, M. R.; Lambert, C.; Wandlowski, T. *J. Am. Chem. Soc.* **2013**, *135*, 12228.

(35) Xu, B.; Tao, N. J. *Science* **2003**, *301*, 1221.

(36) Wang, C.; Batsanov, A. S.; Bryce, M. R.; Martín, S.; Nichols, R. J.; Higgins, S. J.; García-Suárez, V. M.; Lambert, C. J. *J. Am. Chem. Soc.* **2009**, *131*, 15647.

(37) Tao, N. J. *Nat. Nano.* **2006**, *1*, 173.

(38) Segal, D.; Nitzan, A.; Hänggi, P. *J. Chem. Phys* **2003**, *119*, 6840.

(39) Claughton, N. R.; Leadbeater, M.; Lambert, C. J. *J. Phys.: Condens. Mat.* **1995**, *7*, 8757.

(40) Sedghi, G.; García-Suárez, V. M.; Esdaile, L. J.; Anderson, H. L.; Lambert, C. J.; Martín, S.; Bethell, D.; Higgins, S. J.; Elliott, M.; Bennett, N. *Nat. Nanotechnol.* **2011**, *6*, 517.

(41) Zhao, X.; Huang, C.; Gulcur, M.; Batsanov, A. S.; Baghernejad, M.; Hong, W.; Bryce, M. R.; Wandlowski, T. *Chemistry of Materials* **2013**, *25*, 4340.

## Graphical Abstract

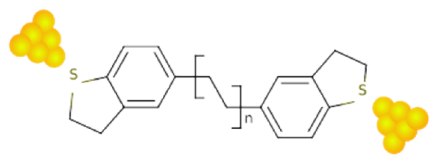

FOR THERMAL MANAGEMENT

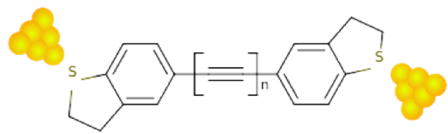

FOR POWER GENERATION

*Supplementary information*

Oligoynes molecular junctions for efficient room temperature thermoelectric power generation

Hatef Sadeghi[*], Sara Sangtarash, and Colin J Lambert[*]

Quantum Technology Centre, Lancaster University, LA1 4YB Lancaster, UK

h.sadeghi@lancaster.ac.uk, c.lambert@lancaster.ac.uk

## 1- Alkane and Oligoyne molecular structure with BT anchor

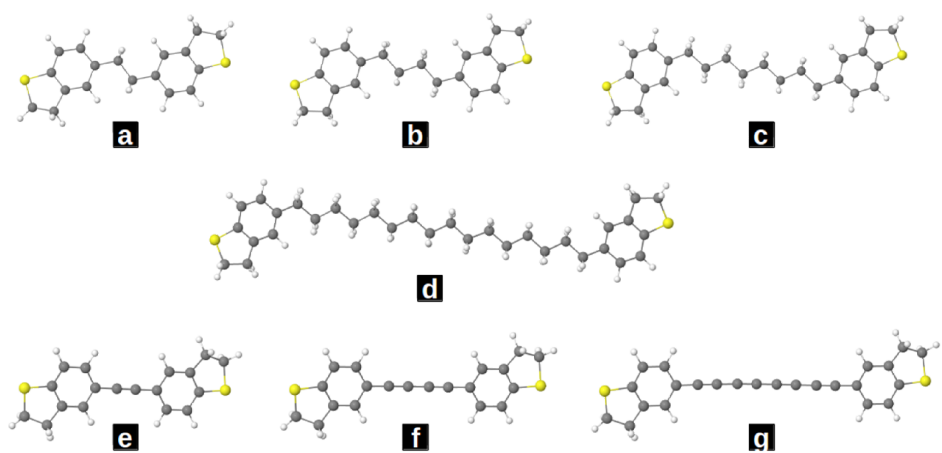

Figure S1. Molecular structure of (a-d) alkanes with $N$ = 1, 2, 4, 8 and (e-f) oligoynes with $N$ = 1, 2, 4.

## 2- Molecule in the junction

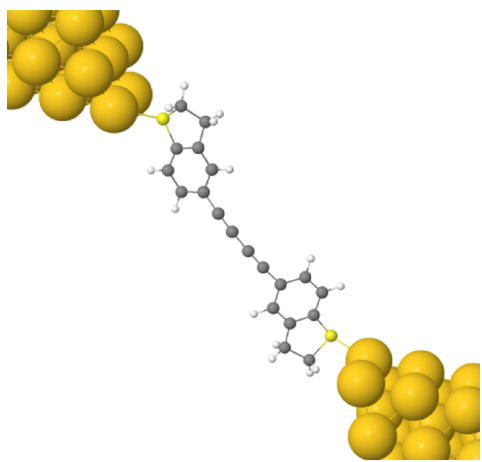

Figure S2. An example of molecule (ie oligoyne with length $N$ = 2 with BT anchor) in a junction between two gold electrodes. The Sulfur atoms in BT anchors are connected to the gold electrodes.

## 3- Tight-binding model calculation

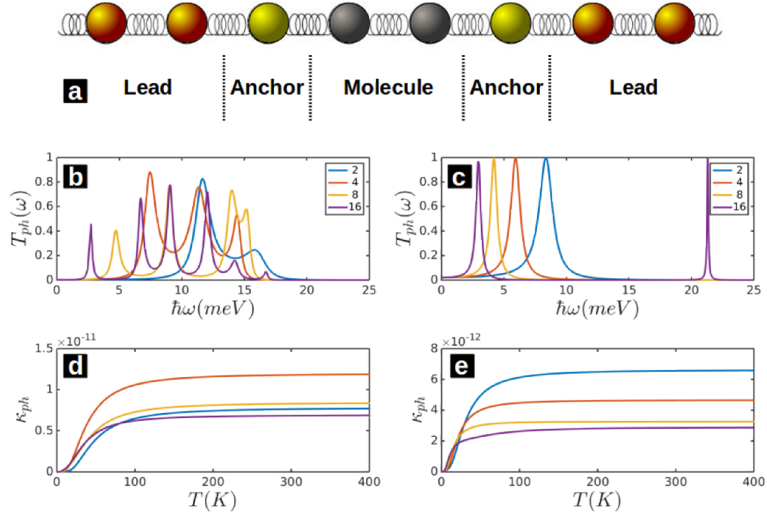

Figure S3. A calculation of the phonon transmission coefficient and thermal conductance of a 1D chain of atoms (with a single degree of freedom) connected by harmonic springs. (a) schematic of junction structure, (b,c) the phonon transmission coefficient of chains of varying length with (b) weak and (c) stronger springs between the atoms and with a fixed coupling between the anchor atom and lead. (d, e) the corresponding thermal conductances (d) for (b) and (e) for (c). In both cases, the resonance widths decrease with increasing length. For the more floppy molecule (b) increasing from $N=2$ to $N=4$ causes more resonances enter the non-filtered low-energy window, leading to an increase in thermal conductance.

## 4- Thermal properties of the alkanes within a junction with higher-Debye-frequency electrodes such as Graphene

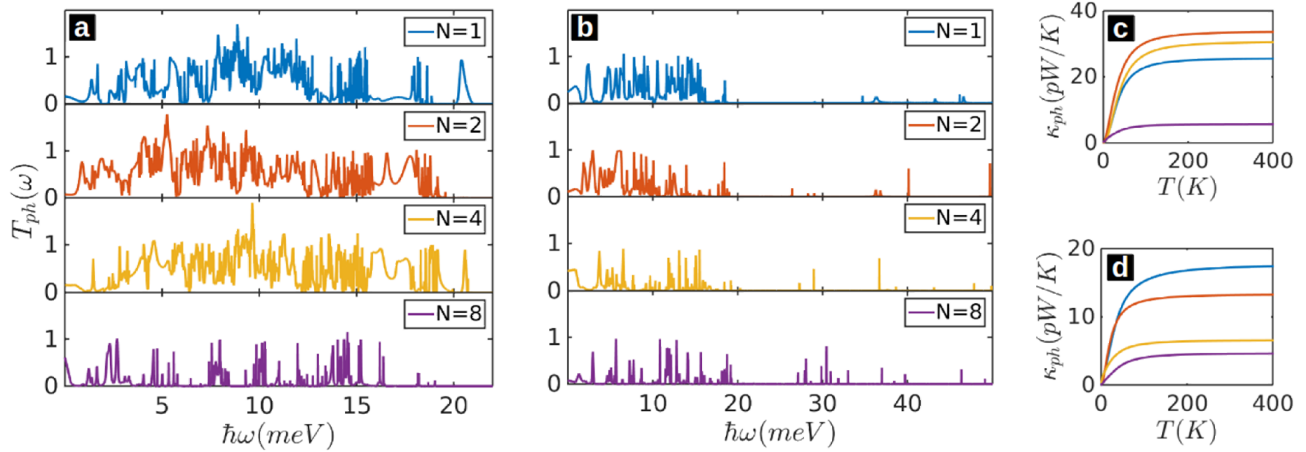

Figure S4. Alkane chains connected to two gold lead with (a, c) normal gold mass and (b, d) reduced gold mass. (a, b) show the phonon transmission coefficient and (c, d) phonon thermal conductance. At high frequencies, the transmission is limited by the number of open phonon channels in the gold electrodes, which falls to zero above the gold Debye frequency. This shows that a molecule that better transmits higher modes, combined with a low-Debye-frequency electrode that filters high energy phonons could be a viable strategy to suppress phonon transmission through the junction.

## 5- Number of open phononic channels in the Gold electrodes

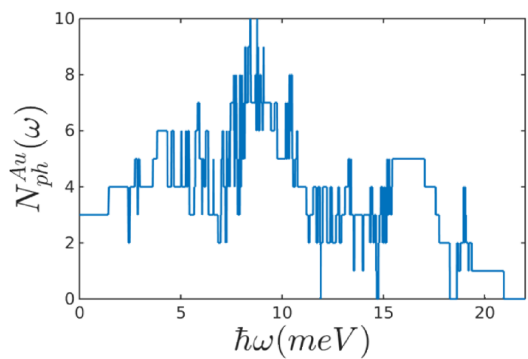

Figure S5. Number of open phononic channels in gold electrodes.

## 6- Alternative anchor groups and electrode tip configurations

We have carried out DFT calculations on the $N=4$ alkane and $N=4$ oligoyne with two new anchors SH, $NH_2$ and with the BT anchor but with different (ie pyramidal) electrode surface, as shown in figure S6. Electronic and phononic transmission coefficients and thermal conductances are then calculated as shown in figures S7-S9.

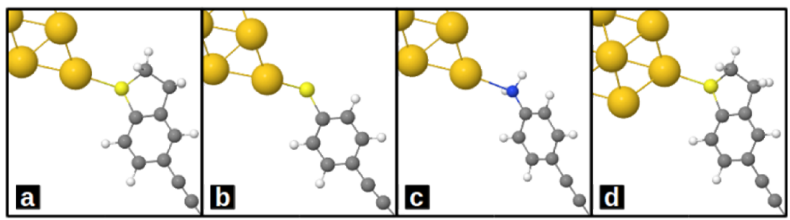

Figure S6. Junction configurations. Three different anchors (a) dihydrobenzo[b]thiophene (*BT*), (b) thiol (*SH*) and (c) amine (*NH₂*) anchors connected to the similar electrode and (d) *BT* anchor connected to the electrode with different surface structure (pyramid tip).

## 7- Electronic and phononic properties of the junctions shown in fig S6

Transport properties for the junctions shown in figure S6 are demonstrated below. The main conclusion is the thermal conductance of oligoynes is lower than that of the corresponding alkanes.

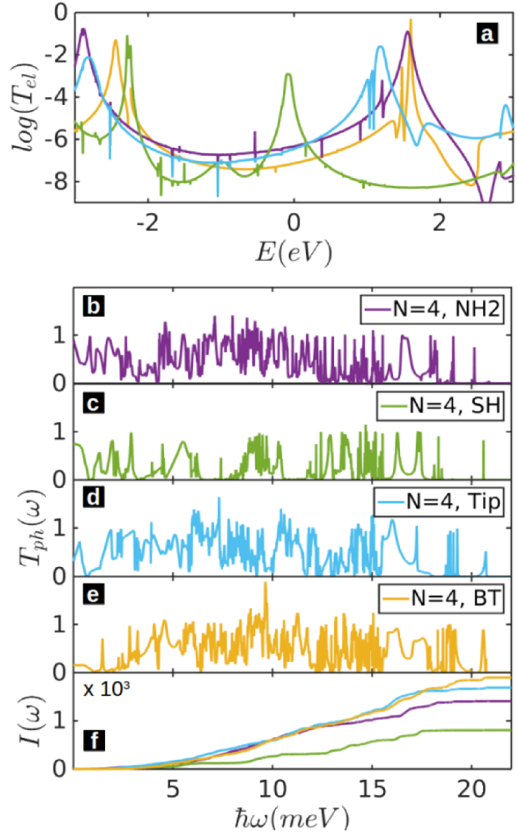
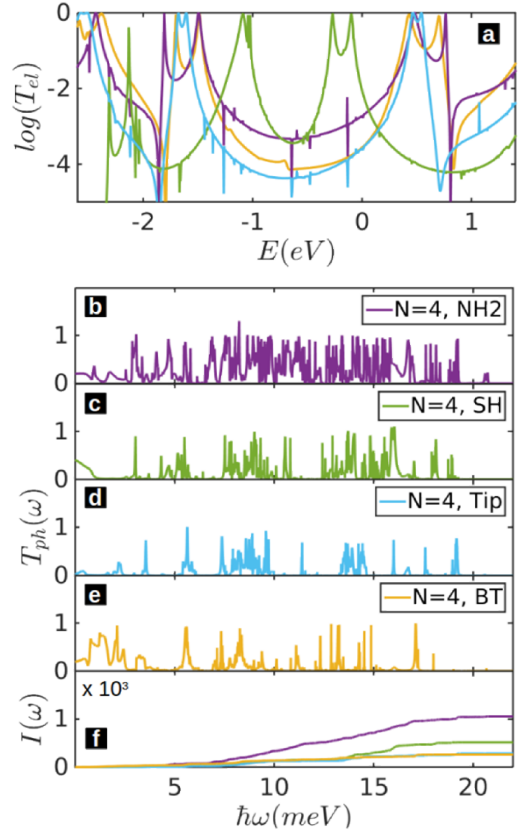

Figure S7. Electronic and phononic properties of alkane $N=4$ junctions shown in fig S6. (a) electronic, (b-e) phononic transmission coefficients, (f) cumulative phononic transmission coefficient $I(\omega)$.

Figure S8. Electronic and phononic properties of oligoyne $N=4$ junctions shown in fig S6. (a) electronic, (b-e) phononic transmission coefficients, (f) cumulative phononic transmission coefficient $I(\omega)$.

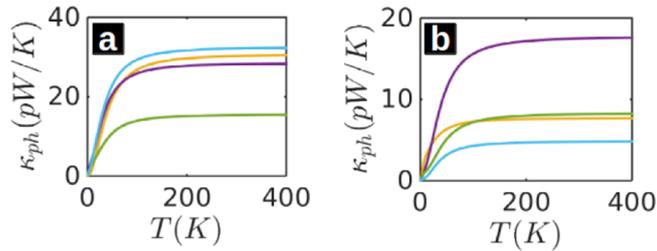

Figure S9. Phononic thermal conductance of the alkanes and oligoynes with different junction structure (shown in fig S6) versus different temperatures. (a) alkanes and (b) oligoynes. Colour map is as figures S7 and S8.

## 8- Electronic and phononic properties of the junctions in the main manuscript with artificially-scaled masses of atoms on the BT anchors or the electrode surface

As a first step towards examining the dependence of phonon transport on tip structure, we have examined the effect of artificially changing the mass of the atoms in the BT anchors by a factor of 0.5 or 2 and of the atoms on the surface of the electrodes by a factor of 2. For both alkanes and oligoynes, the figures below show the effect of decreasing the mass of the anchor atoms by a factor of 0.5 ("light anchor") or of increasing their mass by a factor of 2 ("heavy anchor"). For comparison, the "normal" results obtained without such artificial factors are also shown ("normal"). In all cases, the alkanes show an initial rise in $\kappa_{ph}$ upon increasing the length from $N$=1 to $N$=2. Furthermore, the thermal conductance of oligoynes is lower than that of the corresponding alkanes.

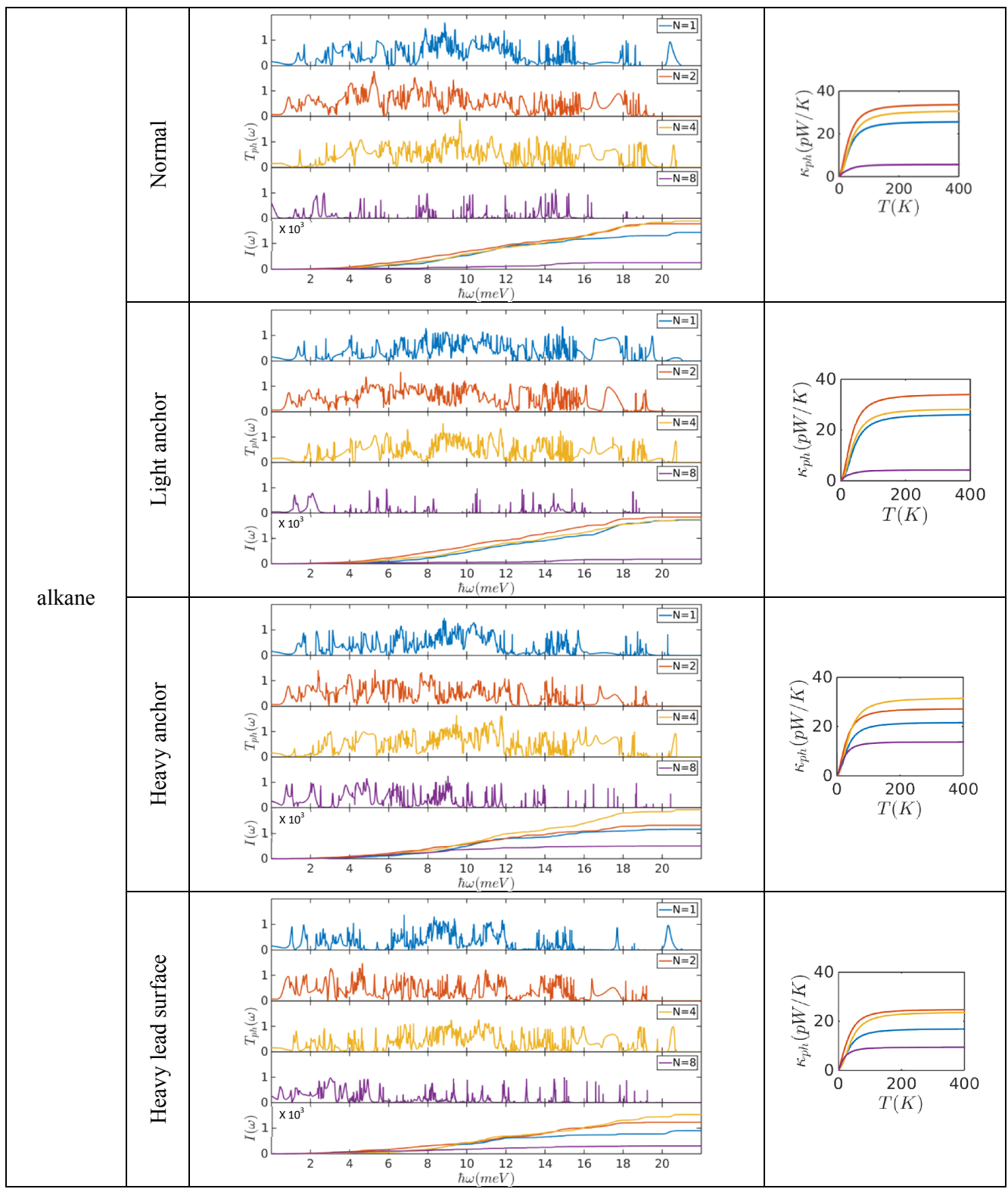

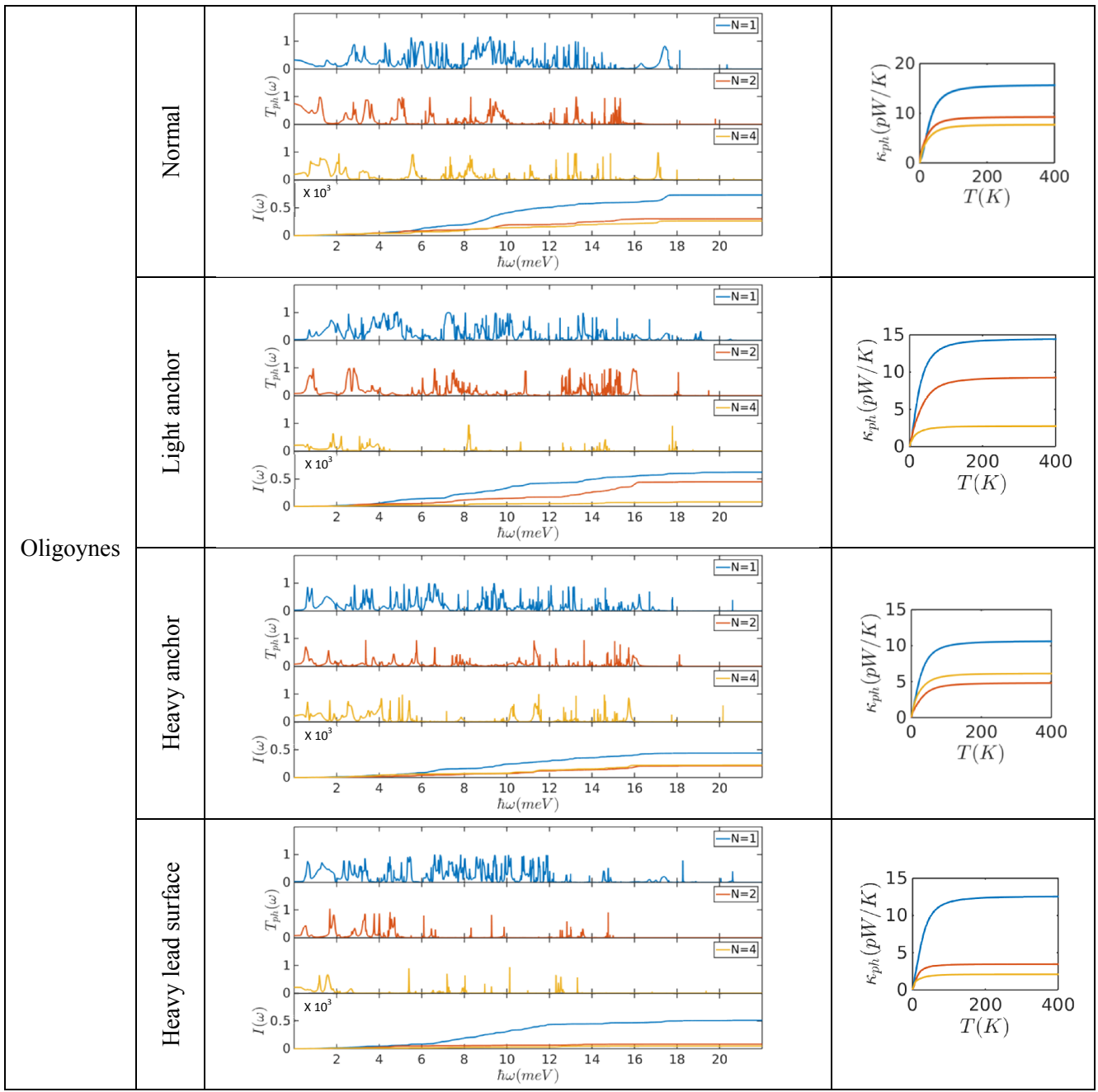

Figure S10. Thermal conductance of the alkanes and oligoynes with artificially-scaled masses of atoms on the BT anchors or the electrode surface.

6